\documentclass[review]{elsarticle}

\usepackage{lineno,hyperref}
\modulolinenumbers[5]

\journal{Chemical Engineering Science}

\usepackage{color}
\usepackage{multirow}
\usepackage{dcolumn}% Align table columns on decimal point
\usepackage{multicol}
\usepackage{graphicx}
\usepackage{caption}
\usepackage{amssymb}
\usepackage{amsmath}
\usepackage{epstopdf}
\usepackage{mathtools}
\usepackage{float}
\usepackage{bm}% bold math

\usepackage{subcaption}
\usepackage{graphicx}
\usepackage{soul}

\usepackage{color}

\biboptions{authoryear}

\biboptions{sort&compress}
\usepackage{color}

\def\i{{\rm i}}
 \def\v{{\bf v}}

\newcommand{\ks}{\textcolor{black}} % for commenting

\begin{document}

\begin{frontmatter}

\title{Exacerbation of viscoelastic instability due to viscous heating}
\author{Ankush Kamboj$^a$, Ramkarn Patne$^b$, P. A. L. Narayana$^a$ and \\ Kirti Chandra Sahu$^b$\footnote{ramkarn@che.iith.ac.in, ananth@math.iith.ac.in, ksahu@che.iith.ac.in}}
\address{$^a$Department of Mathematics, Indian Institute of Technology Hyderabad, Kandi - 502 284, Telangana, India \\
$^b$Department of Chemical Engineering, Indian Institute of Technology Hyderabad, Kandi - 502 284, Sangareddy, Telangana, India}

\begin{abstract}
The linear stability analysis of the pressure-driven flow of an Oldroyd-B fluid through a plane channel is performed to examine the effects of viscous heating-induced buoyancy on the ``purely elastic instability" predicted by \citet{khalid2021continuous} (Phys. Rev. Lett., 2021, 127, 134502). We do not impose any external heating; rather, the temperature increase in the system is solely due to the viscous heating generated by the flow of a highly viscous fluid, which induces buoyancy.   This buoyancy effect adds an extra term to the momentum equation, proportional to the ratio of the Grashof and Reynolds numbers. Since the elastic instability manifests at very low Reynolds numbers, this buoyancy term is crucial for determining flow stability. Our analysis indicates a significant decrease in the critical Weissenberg number due to the viscous heating-induced buoyancy effect, implying that the elastic instability could potentially be observed experimentally at a significantly lower Weissenberg number than that predicted by \citet{khalid2021continuous}. The asymmetry in the streamwise velocity eigenfunctions, resulting from the presence of viscous heating, is found to be the mechanism behind the predicted destabilizing effect.
\end{abstract}

\end{frontmatter}

\clearpage


\section{Introduction}
\label{sec:intro}
Pressure-driven channel flows of highly viscous viscoelastic fluids are commonly encountered in various polymer processing and chemical industries, such as injection moulding, fiber spinning, film blowing, and extrusion \citep{pearson1985mechanics}. For a pressure-driven Newtonian fluid flow through a plane channel, the Tollmien-Schlichting (TS) mode becomes unstable for Reynolds number, $Re (\equiv {U_m H/\nu})$ equal to 5772 \citep{orszag1971accurate}. Here, $H$ represents the half-channel height, $U_m$ denotes the centerline velocity, and $\nu$ is the kinematic viscosity. Due to the high viscosity and low thermal conductivity of polymeric materials, the heat generated by viscous dissipation can cause significant temperature changes. According to \citet{sahu2010stability}, the effect of viscous heating-induced buoyancy on the TS mode is exerted via the term $(Gr/Re) \theta$ appearing in the momentum equations in the Boussinesq framework. Here, $Gr = {\beta_T \Delta T g H^3/\nu^2}$ is the Grashof number, where $\beta_T$ is the volumetric thermal expansion coefficient, $\Delta T$ represents the temperature difference between the centerline and channel walls, and $\theta$ denotes the scaled temperature. The influence of viscous heating on the linear instability characteristics of pressure-driven channel flow has also been examined by \citet{sahu2010numerical,sahu2011jfe,reddy2011effects,srivastava2017temporal}. For the TS mode, since $Re$ is substantially high, the term $\frac{Gr}{Re} \theta$ becomes significant when $Gr \ge Re$. This implies the necessity of an extremely high $Gr$, which may not be achievable in experiments. Note that $Gr$ is independent of velocity; thus, it remains unaffected by the fluid velocity, while the Reynolds number decreases with decreasing fluid velocity. This indicates that the effect of viscous heating-induced buoyancy could become significant if the velocity is very low, implying $Re \ll 1$. Although a Newtonian fluid is linearly stable for $Re \ll 1$, recently, \citet{khalid2021continuous} demonstrated the existence of a new purely elastic mode that persists even for $Re \to 0$ in a highly viscous liquid. In this limit, the term $\frac{Gr}{Re} \theta$ becomes dominant. This naturally leads to the question: can viscous heating-induced buoyancy affect the elastic mode predicted by \citet{khalid2021continuous}? The present study aims to answer this question. Below, we discuss the relevant literature.

\citet{ho1977stability} and \citet{lee1986stability} demonstrated that the plane Poiseuille flow of a viscoelastic fluid is linearly stable at low Reynolds numbers $(Re < 1)$. At sufficiently high $Re$ ($Re > 2000$), employing the upper-convected Maxwell (UCM) model for viscoelastic fluids, Denn and co-workers \citep{porteous1972linear,ho1977stability} identified three new unstable wall modes. Two of these modes were absent in the Newtonian limit, while the third was a continuation of the Tollmien-Schlichting (TS) mode. \citet{sureshkumar1995linear} investigated the linear stability characteristics of plane Poiseuille flow of viscoelastic fluids modeled using the UCM, Oldroyd-B, and Chilcott-Rallison models. For the UCM model, they predicted a non-monotonic behavior of the critical Reynolds number with increasing elasticity number ($E$). Subsequently, \citet{sadanandan2002viscoelastic} examined the stability of plane Poiseuille flow of Oldroyd-B fluid, which also exhibited non-monotonic behavior of the critical Reynolds number with respect to $E$. They explored the mechanism behind this behavior by conducting an analysis of the vorticity and kinetic energy budgets associated with the most dangerous disturbances. Similar non-monotonic behavior was observed by \citet{zhang2013linear} for inertia-dominated channel flow modeled by the FENE-P model. Using linear stability analysis and direct numerical simulations (DNS), \citet{brandi2019dns} studied the effect of elasticity and polymer concentration on the channel flow of Oldroyd-B fluid. They found that the flow becomes more stable as polymer concentration decreases, and the effect of elasticity is dependent on polymer concentration.

It is well known that Hagen–Poiseuille flow of a Newtonian fluid is linearly stable at all Reynolds numbers \citep{meseguer2003linearized}. However, experiments show that the laminar-turbulent transition occurs at a Reynolds number of around 2000 for Newtonian fluid flow through a pipe \citep{avila2011onset}. Experiments by \citet{samanta2013elasto} and \citet{choueiri2018exceeding} demonstrated that adding polymers triggers the transition even at Reynolds numbers below 2000. In sharp contrast to the pipe flow of Newtonian fluids, \citet{garg2018viscoelastic} reported an elasto-inertial center mode instability at $Re = 800$ for the flow of an Oldroyd-B fluid in a pipe. A detailed study on the origin of this pipe flow instability can be found in \citet{chaudhary2021linear}. The linear instability in the UCM limit of plane Poiseuille flow was comprehensively studied by \citet{chaudhary2019elasto}. Along with the antisymmetric unstable TS mode, \citet{chaudhary2019elasto} found a symmetric unstable mode. They discovered that both sinuous and varicose modes were unstable, in contrast to the Newtonian case, where only the sinuous mode is unstable. In viscoelastic channel flow, \citet{page2020exact} identified the origin of elasto-inertial turbulence by computing the finite-amplitude nonlinear travelling wave solution, which exhibits an arrowhead structure of polymer stretching. \citet{dubief2022first} analyzed elasto-inertial turbulence using direct numerical simulations (DNS).

Recently, \citet{khalid2021centre} observed that for sufficiently elastic dilute polymer solutions, a center mode instability similar to that reported by \citet{garg2018viscoelastic} occurs, with a phase speed close to the maximum base state velocity. They demonstrated the absence of instability for $\beta < 0.5$, where $\beta$ is the ratio of solvent to solution viscosity. Subsequently, for a highly elastic ultra-dilute polymer solution, \citet{khalid2021continuous} discovered a purely elastic instability in the plane Poiseuille flow of an Oldroyd-B fluid. They demonstrated the continuous pathway connecting elastic turbulence to elasto-inertial turbulence. In another recent study, \citet{priyadarshi2023new} also found two purely elastic instabilities in the gravity-driven flow of an Oldroyd-B fluid on an inclined plane in the creeping-flow limit. One of these purely elastic instabilities is analogous to the center mode instability of \citet{khalid2021continuous}. On the other hand, researchers \citep{groisman2000elastic,groisman2001efficient,steinberg2021elastic} have experimentally studied elastic turbulence in shearing flows in the inertialess limit. A brief review of instabilities in shearing flows of Oldroyd-B fluids, and viscoelastic fluids in general, can be found in \citet{sanchez2022understanding} and \citet{datta2022perspectives}, respectively.

For rectilinear shear flows of Newtonian fluids, Barletta and co-workers \citep{barletta2010convection,barletta2011onset,barletta2023viscous} studied the effect of viscous heating-induced buoyancy on flow stability. They considered asymmetric thermal boundary conditions, with a constant temperature maintained at the upper wall and an adiabatic boundary condition at the lower wall. Subsequently, \citet{barletta2023dissipation} investigated the linear stability analysis of Couette-like adiabatic flow in a channel with viscous dissipation effects. They found that the most unstable mode has an infinite wavelength. Similarly, \citet{sahu2010stability} considered asymmetrical isothermal temperature conditions at both walls in plane-Poiseuille flow and, by conducting a linear stability analysis, showed that viscous heating can destabilize the flow. The linear stability analysis of Taylor-Couette flow of Oldroyd-B fluid with viscous heating by \citet{al1999influence} revealed the destabilizing influence of viscous heating. Additionally, the most unstable modes were found to be axisymmetric and stationary, which aligns well with the experimental study by \citet{white2000viscous}. \citet{becker2000stability} analyzed the linear stability of creeping plane Couette and Poiseuille flows with temperature-dependent viscosity. In contrast to their findings in Couette flow, stabilizing effects of viscous heating were observed for plane Poiseuille flow at all wavenumbers, and no instabilities were found for the considered set of parameters. \citet{al2000energetic} revealed the stabilizing behavior of viscous heating on the linear stability of Dean flow of an Oldroyd-B fluid. Furthermore, for Peclet numbers greater than $O(10^5)$, they found a new axisymmetric, time-dependent unstable eigenmode. Subsequently, in an experimental study on the stability of torsional flows of highly elastic polymer solutions, \citet{rothstein2001non} demonstrated a stabilizing effect of viscous heating on non-isothermal flow. \citet{olagunju2002effect} studied the linear stability of viscoelastic cone-and-plate flow with viscous heating in the axisymmetric case, revealing that viscous heating has a stabilizing effect on both long-wave and short-wave disturbances. Using linear stability analysis, \citet{hirata2021onset} investigate the thermal instability in the plane Poiseuille flow of weakly elastic fluid with viscous dissipation effects. They found that when the imposed temperature gradient is zero, the instability can triggered with the combined effect of viscous dissipation and fluid elasticity. On the other hand, Researchers have also studied the convective instabilities in the Newtonian \cite{barletta2015thermal} and viscoelastic \cite{hirata2010influence,alves2014effects} fluid flow through a porous media.

As the brief review indicates, most of the earlier studies involving viscous heating with Newtonian and non-Newtonian fluids in different configurations considered an imposed asymmetrical temperature gradient in the flow. However, viscous heating can inherently increase the temperature of systems with highly viscous fluids, such as polymers, lava, etc. The most relevant studies in the context of the present work are those by \citet{khalid2021centre} and \citet{becker2000stability}. While \citet{khalid2021continuous} demonstrated the existence of a purely elastic mode in an isothermal channel flow of a viscoelastic fluid, \citet{becker2000stability} analyzed the effects of viscous heating, albeit in the absence of viscous heating-induced buoyancy, on the stability of pressure-driven viscoelastic flow in the creeping-flow limit. However, as discussed earlier, the buoyancy induced by the temperature gradient becomes significant for very low Reynolds number flows, which remains unexplored. To address this gap, in this work, we investigate the linear stability of the plane Poiseuille flow of an Oldroyd-B fluid with viscous heating-induced buoyancy effects in the creeping flow limit.

The rest of the paper is organized as follows. The problem description, the associated governing equations, the base-state solution, and the linearized stability equations are introduced in \S \ref{sec:form}. The description and validation of the numerical methods used in this study are shown in \S \ref{sec:num}. In \S \ref{sec:dis}, we analyze the linear stability characteristics of the system by examining the dispersion and neutral stability curves to discuss the effects of viscous heating and buoyancy. The major outcomes of the present work are summarized in \S\ref{sec:conc}.

\section{Formulation}
\label{sec:form}

%1
\begin{figure}
 \centering
 \includegraphics[width=0.7\textwidth]{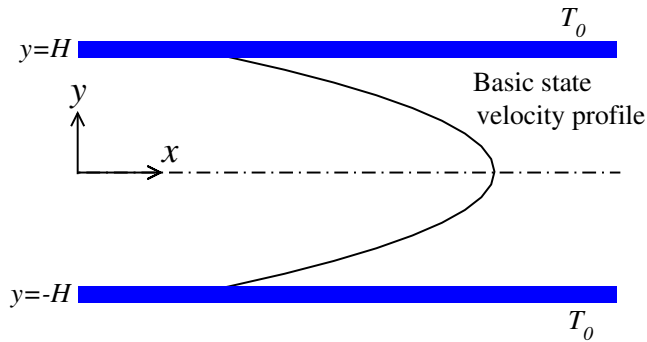}
 \caption{Schematic diagram of the plane-Poiseuille flow of an incompressible Oldroyd-B fluid considered in the present study. The walls at $y = \pm H$ and the fluid are maintained under isothermal conditions at a temperature $T_0$. A typical basic state velocity profile is also depicted in the schematic.}
 \label{fig1}
\end{figure}

As discussed in the introduction, \citet{khalid2021continuous} identified a purely elastic instability in the pressure-driven flow of an incompressible, highly viscous Oldroyd-B fluid within a channel for $Re \ll 1$. It is crucial to highlight that viscous dissipation is inherent in such flows of highly viscous fluids, leading to a rise in fluid temperature even in the absence of external heating, which consequently induces buoyancy effects. Thus, in the present study, we examine the impact of viscous dissipation and buoyancy on the linear stability characteristics of the purely elastic unstable mode observed by \citet{khalid2021continuous}, considering a pressure-driven channel flow of an incompressible Oldroyd-B fluid. To isolate the purely elastic unstable mode from the inertia-driven Tollmien-Schlichting (TS) mode, we investigate the stability characteristics in the low Reynolds number limit. Thus, we fix $Re = 10^{-5}$ unless otherwise specified.

Figure \ref{fig1} illustrates the schematic of the flow configuration considered in the present study. To simplify the analysis, we assume the flow is two-dimensional. A Cartesian coordinate system $(x,y)$ is used, where $x$ and $y$ represent the horizontal and vertical coordinates, respectively, with $y=0$ denoting the channel centerline. The rigid, impermeable channel walls are maintained at a constant temperature $T_0$ and are separated by a distance of $2H$, as shown in figure \ref{fig1}. The flow dynamics is governed by the Cauchy momentum equations and energy equation to incorporate the viscous dissipation along with the constitutive relationship for the Oldroyd-B fluid. We non-dimensionalize these governing equations by employing the half-channel height $(H)$ and the maximum streamwise velocity $(U_m)$ as the length and velocity scales, respectively, while pressure and stresses are scaled by ${\eta U_m/H}$, where $\eta$ denotes the solution viscosity of the Oldroyd-B fluid. The density is given by $\rho = \rho_0\left[ 1 - \beta_T( T - T_0) \right]$, where $\rho_0$ is the reference density and $\beta_T$ is the volumetric thermal expansion coefficient. The dimensionless temperature is defined as $\theta=(T-T_0)/(T_0 - T_c)$, where $T_c$ is the temperature of the fluid at the centerline of the channel. Since the temperature gradient in the channel arises solely from viscous heating, it is reasonable to apply the Boussinesq approximation by assuming the density $\rho$ to be constant. The dimensionless governing equations are given by
\begin{eqnarray}
    \pmb{\nabla.}\textbf{v}&=&0,
    \label{eq6}
    \\
    Re\left[\frac{\partial \textbf{v}}{\partial t} + (\textbf{v}\pmb{.\nabla})\textbf{v}  \right] &=& -\pmb{\nabla} p + \beta\nabla^2\textbf{v} + \pmb{\nabla.}\pmb{\tau} - \frac{Gr}{Re}\theta \pmb{e}_{y},
    \label{eq7}
\end{eqnarray}
where $\v = (u,v)$ is the velocity vector with components $u$ and $v$ in the respective directions, $p$ denotes pressure, and $t$ represents time. In equation (\ref{eq7}), $Re = {U_m H/\nu}$ is the Reynolds number, $Gr = {\beta_T \Delta T g H^3/\nu^2}$ is the Grashof number, $\beta = {\eta_s/\eta}$ is the ratio of solvent to solution viscosity of the Oldroyd-B fluid, $\pmb{\tau}$ is the polymeric stress tensor, and $\pmb{e}_{y}$ is the unit vector in the $y$ direction; $\nu = {\eta/\rho_0}$ is the kinematic viscosity of the polymer solution. The dimensionless form of the Oldroyd-B constitutive relation \citep{larson1988constitutive} for the polymeric stress tensor, $\pmb{\tau}$, is given by
\begin{align}
&\pmb{\tau} + W \left[\frac{\partial \pmb{\tau}}{\partial t}  + (\textbf{v}\pmb{.\nabla})\pmb{\tau} - (\pmb{\nabla} \textbf{v})^T\pmb{.\tau} - \pmb{\tau.}(\pmb{\nabla} \textbf{v}) \right]=     (1 - \beta)(\pmb{\nabla} \textbf{v}   + \pmb{\nabla} \textbf{v}^T),
\label{eq8}
\end{align}
wherein $W= \lambda U_m/H$ is the  Weissenberg number and $\lambda$ is the relaxation time of the fluid. The dimensionless energy equation with viscous dissipation term is given by
\begin{equation}
    Pe\left[\frac{\partial \theta}{\partial t} + (\textbf{v}\pmb{.\nabla})\theta \right]
   = \nabla^2\theta + Br~ (\beta(\pmb{\nabla} \textbf{v} + \pmb{\nabla} \textbf{v}^T) + \pmb{\tau})\pmb{:\nabla}\textbf{v}, 
   \label{eq9}
\end{equation}
where $Pe=\rho_0 c_p U_m H /\kappa$ is the  P\'eclet number and $Br= \eta U_m^2 /\kappa \Delta T$ is the Brinkman number, wherein $c_p$ is the specific heat at constant pressure, $\Delta T=T_0-T_c$, and $\kappa$ is the thermal conductivity.

\subsection{Basic state}
The base state, about which the linear characteristics of the purely elastic mode will be analyzed, corresponds to an undisturbed, steady, and fully developed unidirectional flow. Thus, the base state solutions for the streamwise velocity $(U)$ and temperature field $(\Theta_0)$ obtained from equations (\ref{eq6})–(\ref{eq9}) are given by
\begin{eqnarray}
    U &=& 1 - y^2,
    \label{eq10}
\\
    \Theta_0 &=& \frac{1}{3}Br(1-y^4),
    \label{eq12}
\end{eqnarray}
where $P$ and $\Theta_0$ denote the base state for the pressure and temperature, respectively. The base state of the polymeric stress tensor $\textbf{T} = [T_{ij}]$ is given by
\begin{equation}
    \textbf{T}=\begin{bmatrix}
           2W(1-\beta)(U'^2) & (1-\beta)U'\\
           (1-\beta)U' & 0
\end{bmatrix},
\label{eq11}
\end{equation}
where the prime denotes differentiation with respect to $y$, and the indices $i = (x,y)$ and $j = (x,y)$.

\subsection{Linear Stability Analysis}
We perform a temporal linear stability analysis on the aforementioned base flow by employing a normal mode analysis. Each flow variable decomposes into the base state and a two-dimensional disturbance (symbolized by the hat) as
\begin{eqnarray}
 \left(u,v,p,\theta,\tau_{xx},\tau_{xy},\tau_{yy}\right)\left(x, y, t\right) =  
 \left[U(y),0,P(x),\Theta_0(y), T_{xx}(y),T_{xy}(y),T_{yy}(y)\right] \nonumber   \\
  \hspace*{-4cm} + \left(\hat{u},\hat{v},\hat{p},\hat{\theta},\hat{\tau_{xx}},\hat{\tau_{xy}},\hat{\tau_{yy}} \right)\left(y\right)\exp[\i k(x-ct)],
\label{eq13}
\end{eqnarray}
where $\i = \sqrt{-1}$, $k$, and $c = c_r + \i c_i$ represent the real-valued streamwise wavenumber and complex wave speed of perturbations, respectively. In the temporal stability analysis conducted in this study, $c_i > 0$, $c_i < 0$, and $c_i = 0$ represent the linearly unstable, stable, and neutrally stable modes, respectively, with the growth rate of the disturbance given by $\omega_i = kc_i$. The linear stability equations are derived (after dropping the hat notation) by substituting equation (\ref{eq13}) into equations (\ref{eq6})–(\ref{eq9}) and linearizing around the base state, following a standard approach \citep{schmid2002stability}. They are given by
\begin{eqnarray}
    v^\prime + \i ku&=&0,
     \label{eq14}
\\
        Re\left[\i k(U-c)u + U'v\right] &=& -\i kp + \beta \left[ u^{\prime\prime}-k^2u\right]  + \i k \tau_{xx} + \tau_{yx}^{\prime},
         \label{eq15}
\\
         Re\left[\i k(U-c)v\right] &=& -p^{\prime}+\beta\left[v^{\prime\prime}-k^2v\right] + \i k \tau_{xy} + \tau_{yy}^{\prime} - \frac{Gr}{Re}\theta,
          \label{eq16} \\
          \left[1 + \i kW(U-c)  \right]\tau_{xx} &=& (1-\beta) [ 2\i k u + 4 \i k W^2 U'^2 u     +   2WU\prime u^\prime  \nonumber \\ &-& 4W^2U'U''v] + 2WU'\tau_{xy}, \label{eq17} \\ 
          \left[1 + \i kW(U-c)  \right]\tau_{xy} &=& (1-\beta) [ u' +  \i kv + 2 \i k W^2 U'^2 v - WU''v] \nonumber \\ &+& WU'\tau_{yy}, \\
                  \left[1 + \i kW(U-c)  \right]\tau_{yy} &=& 2(1-\beta)\left[ v' +  \i kWU'v  \right],   \label{eq19} \\
           Pe\left[ (U-c)\i k \theta + \Theta'_0 v  \right] &=& \theta'' - k^2\theta + Br [(2\beta U'+T_{xy})(\i k  v  +  u') + \i k T_{xx}u   \nonumber \\ &+& U' \tau_{xy} + T_{yy}v'].   \label{eq20}
\end{eqnarray}
The solutions of these linear stability equations are obtained by employing the no-slip and isothermal boundary conditions for the velocity and temperature disturbances, respectively. They are given by
\begin{equation}
    \begin{split}
        u=v= \theta=0 \quad \text{at} \quad  y=\pm 1.
         \label{eq21}
    \end{split}
\end{equation}
The equations (\ref{eq14})–(\ref{eq20}), along with the boundary conditions (\ref{eq21}), represent a generalized linear eigenvalue problem $\mathcal{A}\textbf{x} = c \mathcal{B}\textbf{x}$ that needs to be solved to find the eigenvalues $(c)$ and eigenfunctions $\textbf{x}$ for a given set of parameter values $\beta$, $Re$, $Gr$, $Pe$, and $Br$. We observe that the stability equations reduce to those of \citet{khalid2021continuous} when $Gr$, $Br$, and $Pe$ are set to zero.

\begin{figure}
\centering
\includegraphics[width=0.7\textwidth]{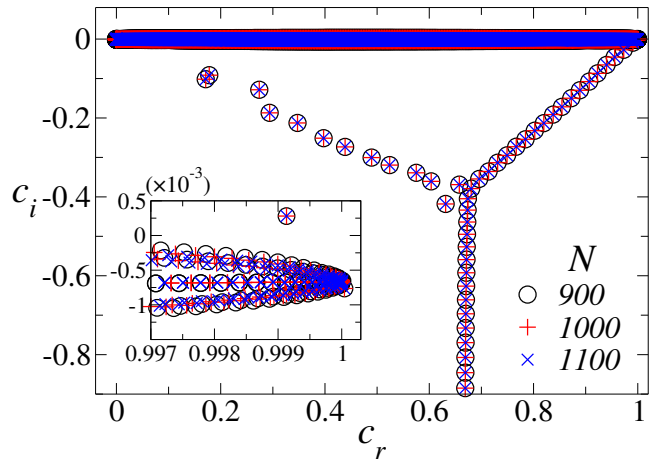}
\caption{Grid convergence test of our numerical scheme showing the effect of the Chebyshev polynomial order $(N)$ on the eigenspectrum. \ks{The axis labels $c_r$ and $c_i$ denote the real and imaginary parts of the eigenvalue $(c = c_r+\i c_i)$, respectively.} The remaining parameter values are: $Re = 10^{-5}$, $Gr = 0.05$, $Pe = 10^{4}$, $Br = 0.05$, $\beta = 0.997$, $W = 2500$, and $k = 0.6$.}
\label{grid con}
\end{figure}

%%%%%%%%%%%%%%%%%%%%%%%%%%%%%%%%%%%%%%%%%%%%%%%%%%%%%%
\section{numerical techniques and validation}
\label{sec:num}
We employ the Chebyshev spectral collocation method \citep{canuto1988spectral} to solve the generalized linear eigenvalue equations, wherein each dynamical variable is expressed as a finite sum of Chebyshev polynomials. We discretize the computational domain along the $y$-axis of the channel using the Gauss-Lobatto collocation points defined as
\begin{equation}
y_j=cos\left(\frac{\pi j}{N}\right), \quad j=0,1,2,...,N.
\label{eq22}
\end{equation}
Here, $N$ represents the order of the base polynomial, with $(N + 1)$ grid points coinciding with all extremums of the Chebyshev polynomial $T_N(y) = \cos(N \cos^{-1} y)$ of order $N$. By discretizing along the $y$-axis, the linear stability equations can be formulated as a generalized matrix eigenvalue problem, expressed as follows:
\begin{align}
\mathcal{A}\textbf{x} = c \mathcal{B}\textbf{x}.
 \label{eq.eig}
\end{align}
The eigenvalue $c$ and eigenfuction $\textbf{x}=[u~ v~ p~ \tau_{xx}~ \tau_{xy}~ \tau_{yy} ~\theta]^T$ of equation (\ref{eq.eig}) is determined using the \textbf{eig} command in \textsc{Matlab}$^{\circledR}$ software. It is important to note that the Chebyshev spectral collocation method may yield some spurious eigenvalues in addition to the physical eigenvalue. We eliminated these unphysical spurious eigenvalues by running the code for different grids by varying the value of $N$ and discarding the eigenvalues that do not meet the tolerance requirement \citep{bewley1998optimal, orszag1971accurate}. To choose the value of $N$, we performed a grid convergence test as shown in figure \ref{grid con}. We plotted the eigenspectrum for different values of $N$ in figure \ref{grid con}. The values of the other parameters in figure \ref{grid con} are $Re = 10^{-5}$, $Gr = 0.05$, $Pe = 10^{4}$, $Br = 0.05$, $\beta = 0.997$, $W = 2500$, and $k = 0.6$, corresponding to a typical set of parameters considered in the present study. The results are obtained for $N=900$, $1000$, and $1100$. It can be seen in figure \ref{grid con} (also see the enlarged inset in the figure) that the eigenspectra are identical for the different values of $N$, indicating the convergence of the numerical scheme. Therefore, we chose $N = 1000$ to plot the rest of the results. Our numerical scheme has also been validated by comparing the results obtained from the present stability solver with previously published results for plane Poiseuille flows of both Newtonian \citep{schmid2002stability} and upper-convected Maxwell (UCM) fluids \citep{sureshkumar1995linear} in Tables \ref{table:1} and \ref{table:2}, respectively. \ks{It can be observed that, while the imaginary part of the eigenvalue shows slight deviation from the results of \citet{sureshkumar1995linear}, it agrees well with the findings of \citet{khalid2021centre}.} It is to be noted that the linear stability equations for a Newtonian fluid can be recovered by setting the values of $Gr$, $Br$, $Pe$, $W$, and $\beta$ to zero in the current formulations (equations \ref{eq14}–\ref{eq20}). Additionally, the stability equations of \citet{sureshkumar1995linear} are obtained by setting $Gr$, $Br$, and $Pe$ to zero. While Table \ref{table:1} shows excellent agreement (ten unstable eigenvalues in descending order) between the results of our numerical procedure and those of \citet{schmid2002stability} for plane Poiseuille flow of Newtonian fluid, Table \ref{table:2} depicts agreement with the results of \citet{sureshkumar1995linear}, who investigated the linear stability of viscoelastic fluid flow modeled using the Oldroyd-B constitutive equation.

\begin{table}
\centering
\begin{tabular}{c c}
\citet{schmid2002stability}     & Present study  \\  \hline
 $0.31210030-0.01979866\i$ & $0.31210030-0.01979866\i$ \\  
 $0.42418427-0.07671992\i$  & $0.42418426-0.07671992\i$ \\  
 $0.92078667-0.07804706\i$  & $0.92078666-0.07804706\i$ \\ 
 $0.92091806-0.07820060\i$ & $0.92091806-0.07820060\i$ \\ 
 $0.85717055-0.13990151\i$ & $0.85717055-0.13990151\i$ \\    
 $0.85758968-0.14031674\i$ & $0.85758968-0.14031674\i$ \\  
  $0.79399812-0.20190508\i$  & $0.79399812-0.20190508\i$ \\  
 $0.79413424-0.20232063\i$  & $0.79413424-0.20232063\i$ \\ 
 $0.63912513-0.22134137\i$  & $0.63912513-0.22134137\i$ \\  
 $0.53442105-0.22356175\i$  & $0.53442105-0.22356174\i$  \\ \hline
\end{tabular}
 \caption{Comparison of the ten leading least stable eigenvalues obtained using the present linear stability solver with those of \citet{schmid2002stability} for plane Poiseuille flow at $Re = 2000$ and $k = 1$. To obtain the Newtonian fluid results of \citet{schmid2002stability}, we set the values of $Gr$, $Br$, $Pe$, $W$, and $\beta$ to zero in our formulation for the Oldroyd-B fluid, which accounts for viscous dissipation and buoyancy effects.}
\label{table:1}
\end{table}

\begin{table}[h]
\centering
\begin{tabular}{c c c c c c c}
 $Re$  & $\beta$  & $k$  & $W$  & \cite{sureshkumar1995linear}  & \citet{khalid2021centre}  & Present study\\ \hline
  $1990$  & 0  & 1.2  & $5.97$  & $0.34580+0.000101 \i$  & $0.34580 +0.0000098 i$  & $0.34580 +0.0000098 \i$\\
  $3960$  & $0.5$  & $1.15$  & $3.96$  & $0.29643 + 0.00000017 \i$  & $0.29643 + 0.00000017 i$  & $0.29643 + 0.00000017 \i$\\ \hline
\end{tabular}
 \caption{\ks{Comparison of the least stable eigenvalue obtained from our numerical solver with the corresponding eigenvalue from \citet{sureshkumar1995linear} and \citet{khalid2021centre} for two sets of parameters. This study investigates the linear stability of a viscoelastic fluid flow modelled using the Oldroyd-B constitutive equation. For this comparison using the current formulation, we set the values of $Gr$, $Br$, and $Pe$ to zero.}}
\label{table:2}
\end{table}

% \begin{table}
% \centering
% \begin{tabular}{c c c c c c c}
%  $Re$ & $\beta$ & $k$ & $W$  & \citet{sureshkumar1995linear} & Present study\\ \hline
%   $1990$ & 0 & 1.2 & $5.97$ & $0.34580+0.000101\i$ & $0.34580 +0.0000098\i$ \\  
%  $3960$ & $0.5$ & $1.15$ & $3.96$ & $0.29643 + 0.00000017\i$  & $0.29643 + 0.00000017\i$\\ \hline
% \end{tabular}
%  \caption{Comparison of the least stable eigenvalue obtained from our numerical solver with the corresponding eigenvalue from \citet{sureshkumar1995linear} for two sets of parameters. This study investigates the linear stability of viscoelastic fluid flow modelled using the Oldroyd-B constitutive equation. For this comparison, we set the values of $Gr$, $Br$, and $Pe$ to zero.}
% \label{table:2}
% \end{table}

\section{Results and discussion}
\label{sec:dis}

\begin{table}
\centering
\begin{tabular}{c c}
 Physical parameters     & Experimental range  \\  \hline
 Relaxation time $(\lambda)$ & $0.001 - 17 s$ \\
Volumetric thermal expansion coefficient $(\beta_{T})$ & $10^{-4} - 10^{-3} K^{-1}$ \\
Solvent viscosity $(\eta_s)$ & $0.1 - 0.324 Pa .~s$ \\
Solution viscosity $(\eta)$ & $0.101 - 10^5 Pa.~s$ \\
Density $(\rho)$ & $1286 - 1303~ Kg/m^{3}$\\
Thermal conductivity $(\kappa)$ & $0.11 - 0.44 W m^{-1}  K^{-1}$  \\ \hline
\end{tabular}
\caption{The values of the physical parameters associated with different viscoelastic fluids considered in the previous studies \citep{jha2020universal, wapperom1998thermodynamics, varshney2017elastic, groisman2000elastic, Sun_2018, bird1987dynamics, jun2011elastic, ngo2016thermal}.}
\label{table:3}
\end{table}

We begin the presentation of our results by establishing the purely elastic unstable mode found by \citet{khalid2021continuous} in the creeping flow limit $(Re \approx 0)$. Subsequently, we examine the effect of viscous dissipation and the resultant buoyancy effect by varying the Brinkman number $(Br)$, Grashof number $(Gr)$, and Weissenberg number $(W)$ on the linear stability characteristics of this purely elastic unstable mode. In order to get a practical relevance of our study, we listed the physical parameters associated with elastic polymeric fluid and the associated viscous heating terms in Table \ref{table:3}. The corresponding values of the dimensionless parameters are also listed in Table \ref{table:4}.

\begin{table}
\centering
\begin{tabular}{c c}
 Dimensionless parameters     & Range  \\  \hline
 $Gr(\equiv\beta_T \Delta T g H^3/ \nu^2)$ & $10^{-8} - 10^{-1} $ \\
 $Re(\equiv U_m H/\nu)$ & $10^{-3} - 1$\\
 $Br(\equiv \eta U_m^2 /\kappa \Delta T)$ & $10^{-9} - 10^{-1}$\\
 $Pe(\equiv \rho_0 c_p U_m H /\kappa)$ & $10^{3} - 10^{6}$\\
 $W(\equiv \lambda U_m/H)$ & $1-10^{4}$\\
 $\beta(\equiv \eta_s/\eta)$ & $0-1$  \\ \hline
\end{tabular}
 \caption{The range of different dimensionless parameters for viscoelastic fluid based on the fluid properties listed in Table \ref{table:3}.}
\label{table:4}
\end{table}

\begin{figure}
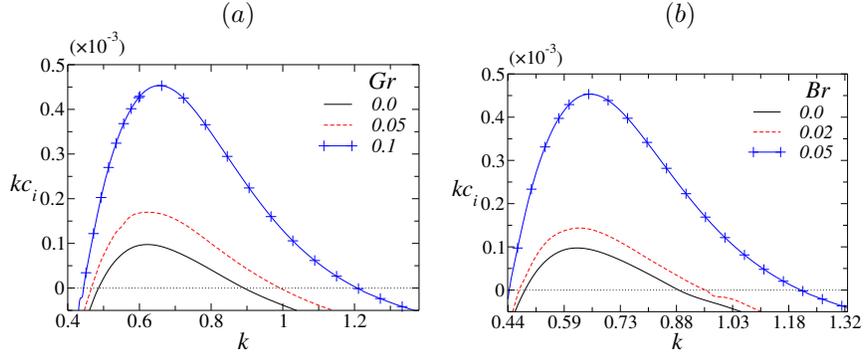

\centering
\hspace{0.6cm}$(a)$ \hspace{5.2cm} $(b)$ \\
\includegraphics[width=0.45\textwidth]{Figure3a.eps}\hspace{3mm} 
\includegraphics[width=0.45\textwidth]{Figure3b.eps} 
\caption{Dispersion curves for different values of $(a)$ Grashof number $(Gr)$ for $Br=0.05$, and $(b)$ Brinkman number $(Br)$ for $Gr=0.1$. The values of the rest of the parameters are $Pe=10^{4}$, $\beta=0.997$, and $W=2500$.}
\label{fig2}
\end{figure}

Figure \ref{fig2}$(a)$ and \ref{fig2}$(b)$ depict the dispersion curves (variations of the growth rate of the disturbance, $kc_i$ versus $k$) for different values of the Grashof number $(Gr)$ for $Br=0.05$ and Brinkman number $(Br)$ for $Gr=0.1$, respectively. The rest of the parameters are $Pe=10^{4}$, $\beta=0.997$, and $W=2500$. The dispersion curves shown in figures \ref{fig2}$(a)$ and \ref{fig2}$(b)$ are paraboloidal, and the growth rate of the disturbance, $kc_i>0$ over a finite band of wavenumbers, indicating the presence of a linear instability; there are also well-defined ``most-dangerous'' and ``cut-off'' modes that correspond to the values of $k$ for which $kc_i$ is maximal and beyond which $kc_i<0$, respectively. It is to be noted that while the Grashof number represents the dimensionless number that measures the ratio of buoyancy to viscous forces in a fluid, the Brinkman number characterizes viscous heating. Thus, the Grashof number becomes relevant in the presence of viscous dissipation (finite Brinkman number). In the situation without viscous dissipation, the system will remain isothermal; thus, the Grashof number becomes irrelevant in the configuration considered in the present study. Therefore, the dispersion curve for $Gr=0$ in figure \ref{fig2}$(a)$ and $Br=0$ in figure \ref{fig2}$(b)$ corresponds to the unstable mode presented by \citet{khalid2021continuous}. It can be seen in figure \ref{fig2}$(a)$ that as we increase the value of \ks{$Gr$}, both the maximum growth rate and the unstable region in the wavenumber $k$ increase, indicating the destabilizing influence of the Grashof number. However, a close inspection of figure \ref{fig2}$(a)$ reveals that the value of the wavenumber associated with the ``most-dangerous'' mode remains approximately the same for different values of \ks{$Gr$} considered. 

%3

Figure \ref{fig2}$(b)$ demonstrates the effect of viscous dissipation on the growth rate of disturbances obtained by varying the values of the Brinkman number with $Gr=0.1$. The rest of the dimensionless parameters $(Re, Pe, \beta, \text{and} W)$ have the same values as considered in figure \ref{fig2}$(a)$. It can be seen in figure \ref{fig2}$(b)$ that increasing $Br$ (increasing the influence of viscous heating) increases the maximum growth rate, which indicates the destabilizing behavior of the viscous dissipation. Moreover, the wavenumber range for the unstable region expands as the $Br$ value increases. Hence, in contrast to the findings of \citet{becker2000stability}, we observe that the viscous heating characterized by the Brinkman number has a destabilizing influence at all wavenumbers. \ks{The discrepancies with \citet{becker2000stability} can be attributed to differences in the governing assumptions. It is to be noted that while they considered viscosity as temperature-dependent and neglected buoyancy effects, our model assumes constant viscosity and incorporates buoyancy. Furthermore, viscous heating in our analysis generates temperature gradients that initiate convection, which is further amplified by buoyancy, an effect not accounted for in \citet{becker2000stability}.} It is also in contrast to the previous studies involving Newtonian fluid in plane Poiseuille and Couette flow configurations, which found that viscous heating is stabilizing \citep{yueh1996linear, wylie2007extensional, sukanek1973stability}. \citet{pinarbasi2005viscous} also observed a similar stabilizing effect of viscous heating in the Poiseuille flow of inelastic fluids. However, \citet{sahu2010stability} and \citet{costa2005viscous} found that viscous heating can trigger unstable flows, albeit they only considered pressure-driven channel flow Newtonian fluid in high Reynolds numbers.

%4
\begin{figure}
\centering
\includegraphics[width=0.9\textwidth]{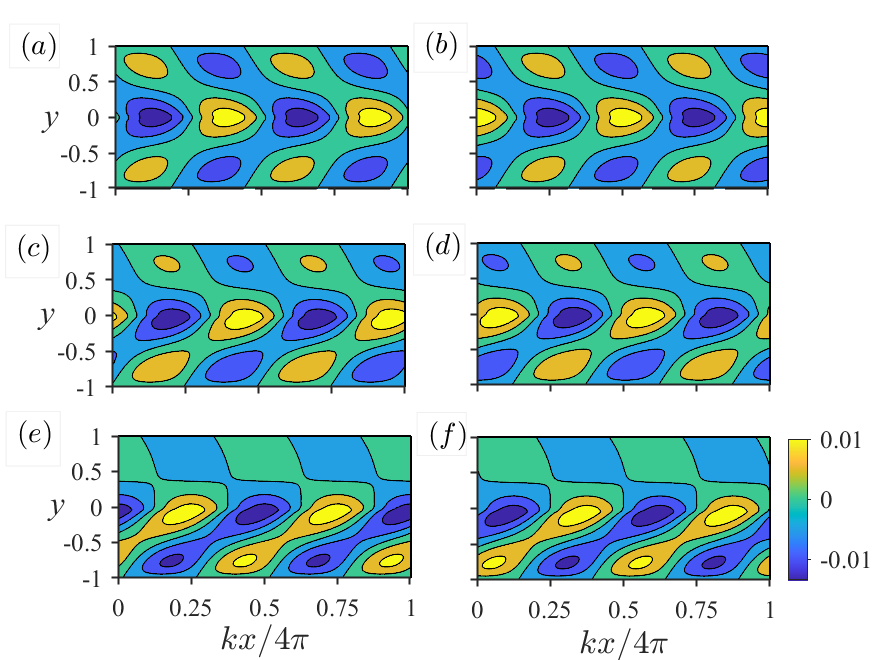} 
\caption{Contours showing the spatial variation of the real $(a,c,e)$ and imaginary $(b,d,f)$ parts of the perturbed streamwise velocity $(\hat{u}(y)\exp[\i k(x-ct)])$ for different values of the Grashof number $(Gr)$ are presented. The first, second, and third rows of this panel correspond to $Gr = 0$, $0.05$, and $0.1$, respectively. The values of the remaining parameters are $Pe=10^{4}$, $k=0.6$, $Br=0.05$, $\beta=0.997$, and $W=2500$. The color map for all the plots is shown at the right end of the bottom panel.}
\label{fig3}
\end{figure}

To gain physical insight into the instability driven by viscous heating, figures \ref{fig3}$(a,c,e)$ and \ref{fig3}$(b,d,f)$ illustrate the effect of increasing the Grashof number $(Gr)$ on the contour plots of the real and imaginary parts of the perturbed streamwise velocity $\hat{u}(y)\exp[\i k(x-ct)]$, respectively. Here we have taken a typical value of $t=1$; a similar demonstration can be provided for other values of $t$. These plots correspond to the unstable mode depicted in figure \ref{fig2}$(a)$ at $k=0.6$. The contours in the first, second, and third rows of this panel are shown for $Gr = 0$, $0.05$, and $0.1$, respectively. For $Gr=0$ (as reported by \citet{khalid2021continuous}), the contours of the real and imaginary parts of the perturbed streamwise velocity exhibit symmetry around the centerline of the channel, as seen in figures \ref{fig3}$(a)$ and \ref{fig3}$(b)$. However, when buoyancy effects are considered by increasing the $Gr$ value to 0.05, the contours in figures \ref{fig3}$(c)$ and \ref{fig3}$(d)$ lose their symmetric characteristics, showing maximum variation near the channel centerline. Further increasing the $Gr$ value, as shown in figures \ref{fig3}$(e)$ and \ref{fig3}$(f)$, results in asymmetric contours, with more pronounced variation in the vicinity of the centerline.

\begin{figure}
\centering
\includegraphics[width=0.9\textwidth]{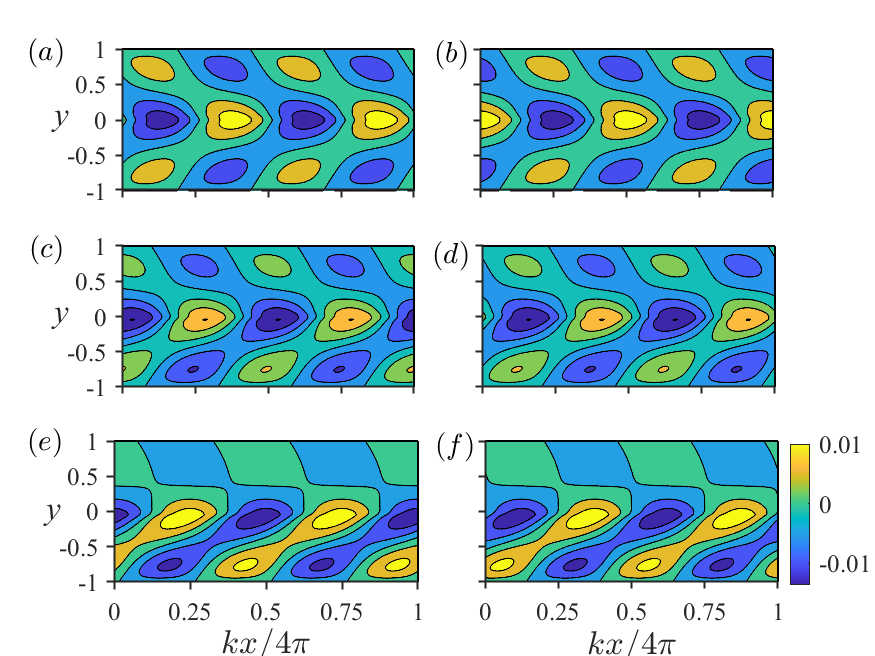} 
\caption{Contours showing the spatial variations of the real $(a,c,e)$ and imaginary $(b,d,f)$ parts of the perturbed streamwise velocity $(\hat{u}(y)\exp[\i k(x-ct)])$ for different values of the Brinkman number $(Br)$ are presented. The first, second, and third rows of this panel correspond to $Br = 0$, $0.02$, and $0.05$, respectively. Here, $t=1$. The values of the remaining parameters are $Pe=10^{4}$, $k=0.6$, $Gr=0.1$, $\beta=0.997$, and $W=2500$. The color map for all the plots is shown at the right end of the bottom panel.}
\label{fig4}
\end{figure}

The contours of the real and imaginary parts of the perturbed streamwise velocity $(\hat{u}(y)\exp[\i k(x-ct)])$ for different values of the Brinkman number $(Br)$ are presented in figure \ref{fig4} for $t=1$. We have plotted figure \ref{fig4} for the unstable mode of figure \ref{fig2}$(b)$ with $k=0.6$. The contours in figures \ref{fig4}$(a)$ and \ref{fig4}$(b)$ at $Br=0$ correspond to the studies of \citet{khalid2021continuous} and exhibit symmetric behavior around the channel centerline. As we consider viscous heating with $Br=0.02$ in figures \ref{fig4}$(c)$ and \ref{fig4}$(d)$, the contours lose their symmetry to some extent, and their spatial variation in the neighborhood of the channel walls is less pronounced compared to the spatial variation at the channel walls in figures \ref{fig4}$(a)$ and \ref{fig4}$(b)$. In figures \ref{fig4}$(e)$ and \ref{fig4}$(f)$, increasing the $Br$ value to $0.05$ results in asymmetric contours with greater variation around the channel centerline and the lower part of the channel.

%5

%6
\begin{figure}
\centering
\includegraphics[width=0.9\textwidth]{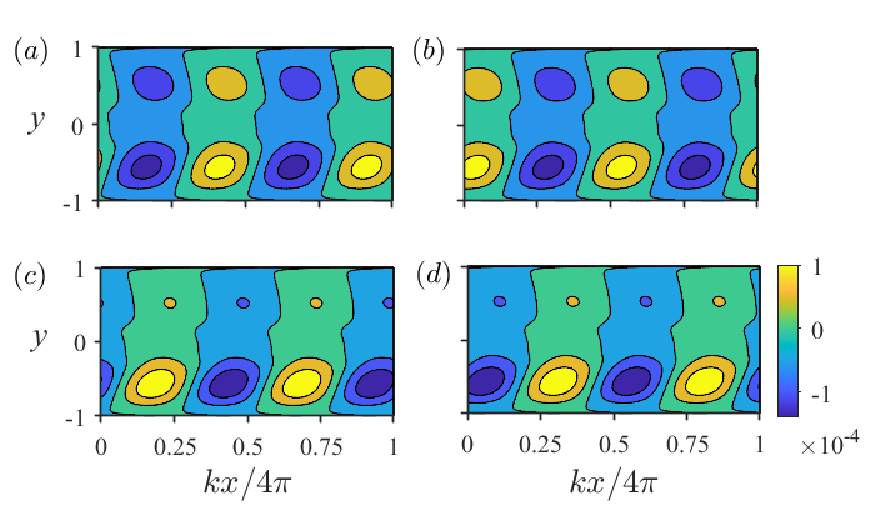} 
\caption{Contours showing the spatial variation of the real $(a,c)$ and imaginary $(b,d)$ parts of the perturbed temperature $(\hat{\theta}(y)\exp[\i k(x-ct)])$ for different values of the Grashof number $(Gr)$ are presented. The first and second rows of this panel correspond to $Gr=0.05$ and $0.1$, respectively. The values of the remaining parameters are $Pe=10^{4}$, $k=0.6$, $Br=0.05$, $\beta=0.997$, and $W=2500$. The color map for all the plots is shown at the right end of the bottom panel.}
\label{fig5}
\end{figure}

%7
\begin{figure*}
\centering
\includegraphics[width=0.9\textwidth]{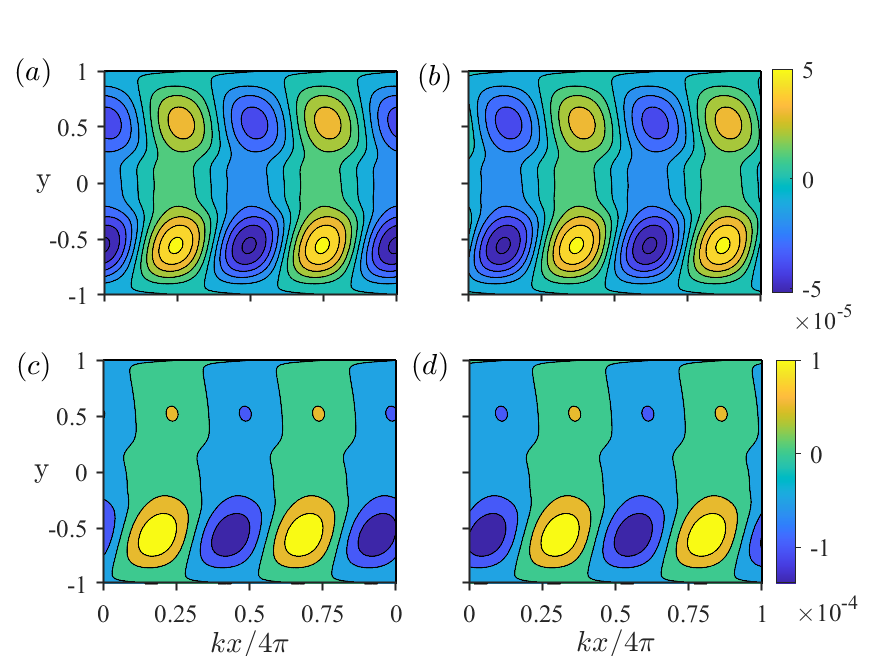} 
\caption{Contours showing the spatial variation of the real $(a,c)$ and imaginary $(b,d)$ parts of the perturbed temperature $(\hat{\theta}(y)\exp[\i k(x-ct)])$ for different values of the Brinkman number $(Br)$ are presented. The first and second rows of this panel correspond to $Br=0.02$ and $0.05$, respectively. The values of the remaining parameters are $Pe=10^{4}$, $k=0.6$, $Gr=0.1$, $\beta=0.997$, and $W=2500$. The color maps for the contours in the first and second rows are shown at their respective right ends.}
\label{fig6}
\end{figure*}

To examine the effect of buoyancy on the perturbed temperature wave $(\hat{\theta}(y)\exp[\i k(x-ct)])$, we present the real and imaginary parts of the perturbed temperature in figures \ref{fig5}$(a,c)$ and \ref{fig5}$(b,d)$ for different values of the Grashof number $(Gr)$. The parameter values are consistent with those in figure \ref{fig3}. In figures \ref{fig5}$(a)$ and \ref{fig5}$(b)$, corresponding to $Gr=0.05$, the contours are concentrated near the channel centerline and the walls. A closer inspection of these figures reveals that fluctuations are more pronounced in the central region of the lower part of the channel compared to the upper part. For $Gr=0.1$, the contours are still concentrated in the lower region of the channel, as shown in figures \ref{fig5}$(c)$ and \ref{fig5}$(d)$. This shift in the perturbed streamwise velocity contours in the lower part of the channel is evident in figures \ref{fig3}$(e)$ and \ref{fig3}$(f)$.

\begin{figure}
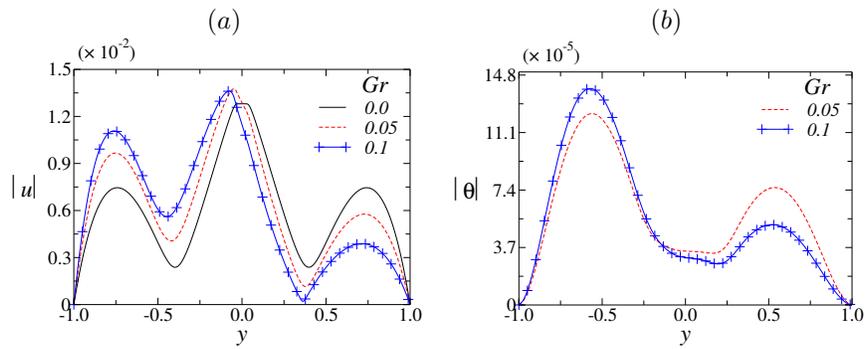

\centering
\hspace{0.6mm} $(a)$ \hspace{5.2cm} $(b)$ \\
\includegraphics[width=0.45\textwidth]{Figure8a.eps}\hspace{3mm} 
\includegraphics[width=0.45\textwidth]{Figure8b.eps} 
\caption{$(a)$ the perturbation in streamwise velocity $u$ and $(b)$ the perturbation in temperature $\theta$ in the $y$-direction for the different values of $Gr$. The values of the rest of the parameters are $Pe=10^{4}$, $Br=0.05$, $k=0.6$, $\beta=0.997$, and $W=2500$.}
\label{fig7}
\end{figure}

\begin{figure*}
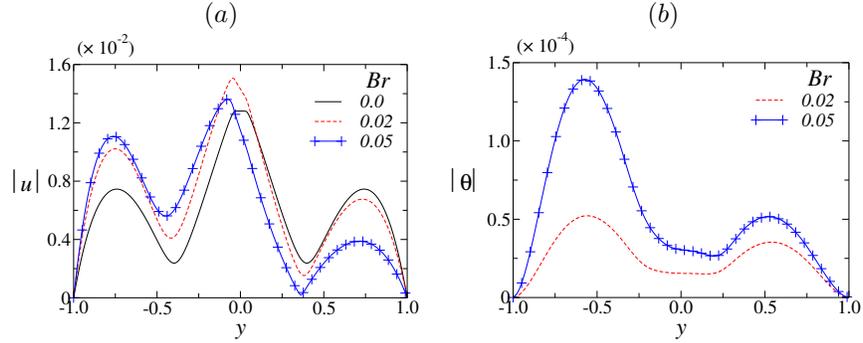

\centering
\hspace{0.6mm} $(a)$ \hspace{5.2cm} $(b)$ \\
\includegraphics[width=0.45\textwidth]{Figure9a.eps}\hspace{3mm} 
\includegraphics[width=0.45\textwidth]{Figure9b.eps} 
\caption{Variations of the absolute values of $(a)$ the perturbation in streamwise velocity $u$ and $(b)$ the perturbation in temperature $\theta$ in the $y$-direction for different values of the Brinkman number $(Br)$. The values of the remaining parameters are $Pe=10^{4}$, $Gr=0.1$, $k=0.6$, $\beta=0.997$, and $W=2500$.}
\label{fig8}
\end{figure*}

We summarise the effect of viscous heating (by varying $Br$) on the real and corresponding imaginary parts of the perturbed temperature $(\hat{\theta}(y)\exp[\i k(x-ct)])$ wave in the $x$-$y$ plane in figure \ref{fig6}$(a,c)$ and figure \ref{fig6}$(b,d)$, respectively. The parameter values are the same as in figure \ref{fig4}. Figures \ref{fig6}$(a)$ and \ref{fig6}$(b)$ illustrate the gradual evolution of the temperature perturbation, which reaches its maximum in the lower part of the channel at $Br=0.02$. As the value of $Br$ increases to $0.05$, as shown in figures \ref{fig6}$(c)$ and \ref{fig6}$(d)$, the magnitude of the temperature perturbation increases, and the fluctuations become more severe. This is because viscous heating intensifies as the $Br$ value increases, leading to a rise in the temperature of the fluid. This explanation also accounts for the destabilizing effect of $Br$ observed in figure \ref{fig2}$(b)$.

Figures \ref{fig7}$(a)$ and \ref{fig7}$(b)$ demonstrate the spatial evolution of the absolute value of the streamwise velocity eigenfunction $(u)$ and temperature eigenfunction $(\theta)$, respectively, corresponding to an unstable mode of figure \ref{fig2}$(a)$ at $k=0.6$. In figure \ref{fig7}$(a)$, it is observed that the curve for $Gr=0$ (corresponds to \citet{khalid2021continuous}) is symmetric about the channel centerline. However, for non-zero values of $Gr$, the curves exhibit asymmetry, as shown in figure \ref{fig7}$(a)$. For all values of $Gr$, the absolute value of the streamwise velocity eigenfunction $(u)$ reaches its maximum near the centerline. As the value of $Gr$ increases in figure \ref{fig7}$(a)$, the absolute value of $(u)$ increases in the lower part of the channel and decreases in the upper part. This trend is mirrored in the curves of the temperature eigenfunction $(\theta)$ in figure \ref{fig7}$(b)$. The aforementioned discussion of figures \ref{fig7}$(a)$ and \ref{fig7}$(b)$ aligns with the outcomes observed in figures \ref{fig3} and \ref{fig5}, respectively. 

Figures \ref{fig8}$(a)$ and \ref{fig8}$(b)$ present the absolute value of the streamwise velocity eigenfunction $(u)$ and temperature eigenfunction $(\theta)$, respectively, corresponding to an unstable mode of figure \ref{fig2}$(b)$ at $k=0.6$. The curves in figure \ref{fig8}$(a)$ are qualitatively similar to those shown in figure \ref{fig7}$(a)$. In figure \ref{fig8}$(b)$, it can be observed that the absolute value of the temperature eigenfunction $(\theta)$ increases as the Brinkman number $(Br)$ increases. This behavior is due to the increase in viscous heating with higher $Br$ values, which was explained in the context of figure \ref{fig6}. Additionally, the findings in figures \ref{fig4} and \ref{fig6} are consistent with the results observed in figures \ref{fig8}$(a)$ and \ref{fig8}$(b)$, respectively.

\begin{figure*}
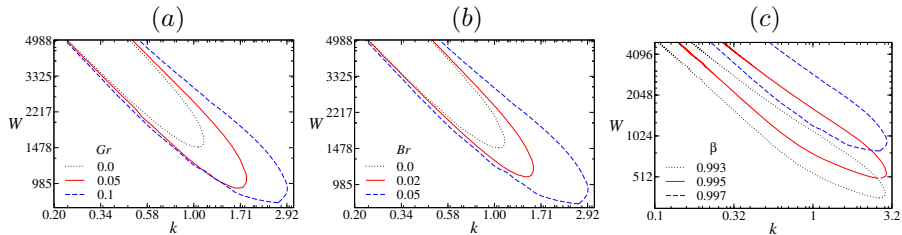

\centering
\hspace{0.2cm} $(a)$ \hspace{3.3cm} $(b)$  \hspace{3.3cm} $(c)$\\
\includegraphics[width=0.32\textwidth]{Figure10a.eps} 
\includegraphics[width=0.32\textwidth]{Figure10b.eps}
\includegraphics[width=0.32\textwidth]{Figure10c.eps}
\caption{Neutral stability curves in the $W-k$ plane for different Grashof number $(Gr)$, Brinkman number $(Br)$, and the ratio of solvent to solution viscosity of the Oldroyd-B fluid $(\beta)$. $(a)$ Effect of the Grashof number $(Gr)$ with $Br=0.05$ and $\beta=0.997$; $(b)$ effect of the Brinkman number $(Br)$ with $Gr=0.1$ and $\beta=0.997$; and $(c)$ effect of $\beta$ with $Gr=0.1$ and $Br=0.05$ on the neutral stability boundaries. Here, $Pe=10^{4}$. These curves illustrate the boundary between stable and unstable regions, with the critical Weissenberg number ($W_c$) and the corresponding wavenumber defining the onset of instability. \ks{The critical Weissenberg numbers ($W_c$) and the corresponding critical wavenumbers ($k_c$) for different values $Gr$, $Br$ and $\beta$ are listed in Tables \ref{T5} - \ref{T7}, respectively.}}
\label{fig9}
\end{figure*}

In figure \ref{fig9}$(a)$, we investigate the impact of buoyancy on the neutral stability curve by varying the Grashof number $(Gr)$ in the $W-k$ plane for $Pe=10^{4}$, $Br=0.05$, and $\beta=0.997$. The neutral stability curves are the loci where $c_i=0$, representing the boundary between stable and unstable regions. The regions inside and outside the neutral stability curve represent unstable and stable zones, respectively. The critical Weissenberg number $(W_c)$ is the minimum value of $W$ on the neutral stability curve, and the corresponding wavenumber is referred to as the critical wavenumber $(k_c)$. The neutral stability curve for $Gr=0$ aligns with the studies by \citet{khalid2021continuous}. From figure \ref{fig9}$(a)$, it is evident that increasing $Gr$ enlarges the unstable region and reduces the critical Weissenberg number, confirming the destabilizing effect of buoyancy. The figure also shows that as $Gr$ increases, both the range of unstable wavenumbers and the critical wavenumber increase. This observation highlights the destabilizing influence of buoyancy on the base flow stability, which validates the conclusions drawn in figure \ref{fig2}$(a)$.

Finally, we examine the critical Weissenberg number $(W_c)$, which characterizes the critical ratio of elastic forces to viscous forces for the pure elastic instability mode, for different values of the Grashof number $(Gr)$, Brinkman number $(Br)$, and the ratio of solvent to solution viscosity of the Oldroyd-B fluid $(\beta)$. The critical Weissenberg number represents the threshold beyond which elastic forces dominate viscous forces, leading to instability in the flow. By varying $Gr$, $Br$, and $\beta$, we explore their influence on $W_c$ to understand how buoyancy, viscous heating, and fluid elasticity contribute to the onset of instability. This analysis reveals the complex interplay between thermal and fluid dynamic effects and provides deeper insight into the stability characteristics of viscoelastic flows under different physical conditions. Figure \ref{fig9} shows the neutral stability curve on the $W-k$ plane for different values of $(a)$ $Gr$ at $Br=0.05$ and $\beta=0.997$, $(b)$ $Br$ at $Gr=0.1$ and $\beta=0.997$, and $(c)$ $\beta$ at $Gr=0.1$ and $Br=0.05$. \ks{The critical Weissenberg numbers ($W_c$) and the corresponding critical wavenumbers ($k_c$) for different values $Gr$, $Br$ and $\beta$ are listed in Tables \ref{T5} - \ref{T7}, respectively.} In all the panels, $Pe=10^4$. The neutral stability curve for $Gr=0$ in figure \ref{fig9}$(a)$ and $Br=0$ in figure \ref{fig9}$(c)$ corresponds to the purely elastic mode observed by \citet{khalid2021continuous}. It can be seen in figures \ref{fig9}$(a)$ and \ref{fig9}$(b)$ that increasing $Gr$ (enhancing the influence of buoyancy) and increasing $Br$ (enhancing the influence of viscous heating) expand the unstable region (region inside the neutral stability curves) and decrease the critical Weissenberg number $(W_c)$ for the onset of instability of the purely elastic mode. This indicates that both the Grashof and Brinkman numbers have a destabilizing effect on the purely elastic mode. It can also be observed in figures \ref{fig9}$(a)$ and \ref{fig9}$(b)$ that the critical wavenumber increases with increasing values of $Gr$ and $Br$. This is in clear contrast to the findings of \citet{becker2000stability}. The outcomes of figure \ref{fig9}$(b)$ corroborate the findings of figure \ref{fig2}$(b)$. 

\begin{table}
\centering
\begin{tabular}{c c c }
 $Gr$ & $W_c$ & $k_c$ \\ \hline
  $0$ & $1490$ & $1.05$ \\ 
 $0.05$ & $935$ & $1.64$\\ 
 $0.1$ & $797.5$ & $2.65$
\end{tabular}
 \caption{\ks{The critical Weissenberg number ($W_c$) and the corresponding critical wavenumber ($k_c$) for different values of $Gr$ associated with Figure \ref{fig9}($a$).}}
\label{T5}
\end{table}

\begin{table}
\centering
\begin{tabular}{c c c }
 $Br$ & $W_c$ & $k_c$ \\ \hline
  $0$ & $1490$ & $1.05$ \\ 
 $0.02$ & $1075$ & $1.45$\\ 
 $0.05$ & $797.5$ & $2.65$
\end{tabular}
 \caption{\ks{The critical Weissenberg number ($W_c$) and the corresponding critical wavenumber ($k_c$) for different values of $Br$ associated with Figure \ref{fig9}($b$).}}
\label{T6}
\end{table}

\begin{table}
\centering
\begin{tabular}{c c c }
 $\beta$ & $W_c$ & $k_c$ \\ \hline
  $0.993$ & $358$ & $2.63$ \\
 $0.995$ & $499$ & $2.59$\\ 
 $0.997$ & $797.5$ & $2.65$ 
\end{tabular}
 \caption{\ks{The critical Weissenberg number ($W_c$) and the corresponding critical wavenumber ($k_c$) for different values of $\beta$ associated with Figure \ref{fig9}($c$).}}
\label{T7}
\end{table}

Inspection of figure \ref{fig9}$(c)$ reveals a qualitative resemblance to that reported by \citet{khalid2021continuous}. The critical Weissenberg number increases with an increase in $\beta$ in figure \ref{fig9}$(c)$. An analogous observation was reported by \citet{khalid2021continuous} (see their figure $2(a)$). At a fixed value of $\beta$, the critical Weissenberg number for the present study is lower than the critical Weissenberg number reported by \citet{khalid2021continuous} in the creeping flow limit. This decrease is attributed to the presence of buoyancy and viscous heating effects, which destabilize the flow. In figure \ref{fig9}$(c)$, we observe a stabilizing effect of $\beta$.

\section{Concluding remarks}
\label{sec:conc}

We have conducted a numerical investigation to study the effect of viscous heating-induced buoyancy on the linear stability characteristics of plane Poiseuille flow of an Oldroyd-B fluid in the low Reynolds number limit. The present investigation is motivated by the purely elastic instability predicted by \citet{khalid2021continuous} in a similar flow configuration of ultra-dilute polymer solutions. Owing to the high Weissenberg number requirement for the purely elastic instability mode to exist, viscous heating will be substantial in such flows. The viscous heating causes an inherent temperature gradient, which in turn induces a buoyancy effect; this effect is more prominent in the low Reynolds number limit. Thus, the buoyancy effect, modeled by employing the term $\frac{Gr}{Re} \theta$ in the momentum equation, becomes crucial when $Gr \ge Re$. Our linear stability results predict a drastic decrease in the critical Weissenberg number compared to the values predicted by \citet{khalid2021continuous}. Hence, viscous heating-induced buoyancy destabilizes the instability predicted by \citet{khalid2021continuous}. Due to viscous heating, the streamwise velocity eigenfunctions lose their symmetry about the center of the channel. Thus, the present study demonstrates that the instability predicted by \citet{khalid2021continuous} could be observed in experiments at a much lower Weissenberg number, indicating that the viscous heating and the associated buoyancy effect should not be neglected in analyzing the stability characteristics of highly viscous viscoelastic fluids. This aspect requires experimental verification. \ks{To experimentally verify the instability predicted by \citet{khalid2021continuous} at lower Weissenberg numbers, influenced by viscous heating and the resulting buoyancy, one could use a dilute solution of high molecular weight polyacrylamide in a viscous sugar syrup, as demonstrated in previous studies \cite{groisman2000elastic}. A pressure-driven channel flow setup with well-insulated walls and active thermal control, such as maintaining constant wall temperature using a coolant, would help isolate the effects of viscous heating. Flow visualization techniques like particle image velocimetry (PIV) combined with infrared thermography could be employed to simultaneously capture flow structures and temperature fields.}

\vspace{2mm}
\noindent{\bf Acknowledgement:}
A.K. acknowledges his gratitude to the University Grants Commission (UGC), India, for providing financial assistance. K.C.S. thanks the Indian Institute of Technology Hyderabad, India for financial support through grants IITH/CHE/F011/SOCH1.

%\vspace{2mm}
%
%\noindent{\bf Author Contributions}
%S. C.: Investigation (equal); Methodology  (equal); Writing – original draft (equal). S. S. A.: Investigation (equal); Methodology  (equal); Writing – original draft (equal).  L. D. C.: Conceptualization; Supervision; Writing – review \& editing. K. C. S.: Conceptualization; Supervision; Writing – review \& editing.

%\section*{Data Availability Statement}
%The data that support the findings of this study are available from the corresponding author upon reasonable request.

\vspace{2mm}

%\nocite{*}
%\bibliography{bibl}% Produces the bibliography via BibTeX.

\end{document}